\documentclass[prl,amsmath,aps,10pt,superscriptaddress,letterpaper,tightenlines,twocolumn]{revtex4}
\usepackage{graphicx,xspace,units}
\usepackage{float}
\usepackage{amsmath}
\usepackage{amssymb}
\usepackage[normalem]{ulem}
\usepackage{wrapfig}
\usepackage{epsfig}
\usepackage[usenames]{color}

\definecolor{MidnightBlue}{cmyk}{0.98,0.13,0,0.43}
\definecolor{DarkGreen}{rgb}{0,0.7,0.1}

\def\a{s}
 \def\b{s}

\newcommand{\add}[1]{\if\a\b{{\color{red} #1}}\else{#1}\fi}
\newcommand{\del}[1]{{\if\a\b{{\color{black}[[#1]]}}\else{}\fi}}

\newcommand{\refeq}[1]{{(\ref{eq:#1})}}
\newcommand{\refeqn}[1]{{Eq.~(\ref{eq:#1})}}
\newcommand{\labeleqn}[1]{\label{eq:#1}}
\newcommand{\reffig}[1]{{Fig.~\ref{fig:#1}}}
\newcommand{\be}{\begin{equation}}
\newcommand{\ee}{\end{equation}}

\newcommand{\ep}{\epsilon}

\newcommand{\half}{\frac{1}{2}}

\begin{document}

\title{Casimir spring and compass: Stable levitation and alignment of compact objects}

\author{Sahand Jamal Rahi}
\email{sjrahi@mit.edu}
\affiliation{Department of Physics,
Massachusetts Institute of Technology,
Cambridge, Massachusetts 02139, USA}

\author{Saad Zaheer}
\affiliation{
Department of Physics and Astronomy,
University of Pennsylvania,
Philadelphia, Pennsylvania 19104, USA}

\begin{abstract}
We investigate a stable Casimir force configuration consisting of an object contained inside a spherical or spheroidal cavity filled with a dielectric medium. The spring constant for displacements from the center of the cavity and the dependence of the energy on the relative orientations of the inner object and the cavity walls are computed. We find that the stability of the force equilibrium can be predicted based on the sign of the force, but the torque cannot be.
\end{abstract}
\maketitle



The Casimir force between atoms or macroscopic objects arises from quantum fluctuations of the electrodynamic field \cite{Casimir48-1,Casimir48-2}. In all known examples it is found that the force is attractive, as long as the space between the objects is empty, and the magnetic susceptibility of the objects is negligible compared to their electric susceptibilities. But when space is filled with a medium with electric permittivity $\ep_M$ intermediate between that of two objects, $\ep_1<\ep_M<\ep_2$, the force between the two becomes repulsive \cite{Dzyaloshinskii61}. This effect has recently been verified experimentally \cite{Munday09}.
But while repulsive forces are nothing new, they become interesting for applications when they produce stable equilibria, which, for example, the Coulomb force cannot. 

For an infinite cylinder enclosed in another, the Casimir force has recently been shown to have a stable equilibrium in the two directions perpendicular to the cylinder axes, when the material properties are chosen so that the force 
between two slabs of the same materials would be repulsive \cite{Rodriguez08-2}.
If the inner and outer cylinders have square cross sections, the direction of the torque exerted by one cylinder on the other is found to agree qualitatively with the predictions of the pairwise summation or proximity force approximations (PSA or PFA). Orientation dependence of the Casimir energy has also been studied recently for a small spheroid facing another spheroid or an infinite plate \cite{Emig09}, or for a small spheroid located inside a spherical metal cavity \cite{Zaheer09}. (The smallness of the spheroid refers to keeping only the first term in the series expansion of the Casimir energy in the largest length scale of the spheroid.) The preferred orientation of the small spheroid flips when it is moved between the inside and the outside of the spherical shell; there is no torque if the spheroid faces an infinite metal plate, which is the limit of an infinitely large shell.

\begin{figure}[ht]
\includegraphics[trim = 0mm 0mm 0mm 0mm,clip,width=8cm]{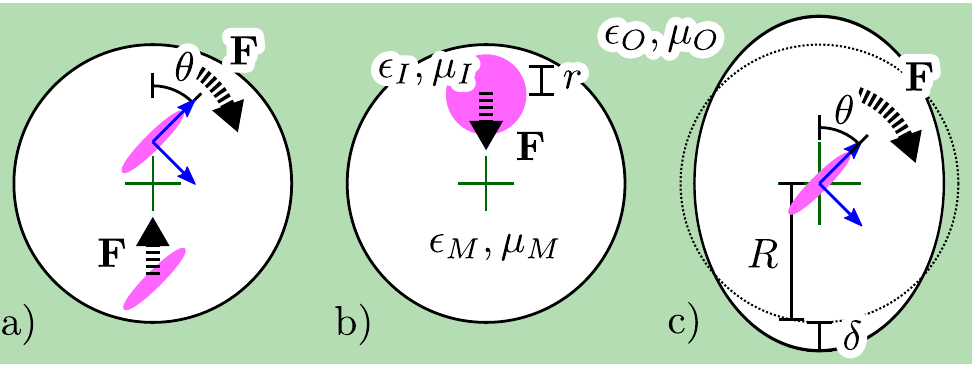}
\caption{(Color online) Summary of the configurations we consider and of the results. To be concrete, we have assumed that the small spheroid's zero frequency permittivity satisfies $\ep_{I,0}>\ep_{M,0}$ and that it is larger in the body-fixed $\hat{\mathbf{z}}$ direction, so $\alpha^E_{zz}>\alpha^E_{\perp\perp}$. Furthermore, the magnetic permeabilities are all set to one. a) Direction of the force on such a spheroid in a spherical cavity if $f_1^E>0$, which holds when $\ep_{M,0}>\ep_{O,0}$, and the direction of the torque when $f_2^E>0$, which is satisfied when either $\ep_{M,0}>\ep_{O,0}$ or $\ep_{M,0}\ll\ep_{O,0}$. b) A finite size sphere experiences a restoring force for the various combinations of materials listed in Table~\ref{table1}. c) Direction of the torque in the center of a slightly spheroidal cavity if $g^E>0$, which requires either $\ep_{M,0}<\ep_{O,0}$ or $\ep_{M,0}\gg\ep_{O,0}$.}
\vspace{-0.3cm}
\label{fig:illus}
\end{figure}

In this paper we investigate the first configuration, depicted in \reffig{illus}, in which a compact object levitates stably due to the Casimir force alone. We characterize the Casimir energy
\be
\mathcal{E} = \mathcal{E}_0 + \half k \tfrac{a^2}{R^2} + \frac{1}{3!} k_3 \tfrac{a^3}{R^3} + \frac{1}{4!} k_4 \tfrac{a^4}{R^4} + \cdots
\labeleqn{Eexp}
\ee
by the spring constant $k$ and the coefficients $k_n$ in a series expansion in $a/R$, where $a$ is the magnitude of the displacement from the center of the cavity and $R$ the radius of the (undeformed) spherical cavity. ($k$ has units of energy here.) Unlike the case of infinite cylinders, where one object can be displaced along the cylinder axes without changing the energy, our case exhibits, for appropriately chosen materials, true stability in all directions and applies to realistic situations. For example, we compute the force on a metal sphere in a spherical drop of liquid surrounded by air. By measuring the mean square deviation from the center, $\langle a^2 \rangle = 3 k_B T R^2/k$, the spring constant $k$ can be verified experimentally. We can estimate that the size of the droplet $R$ has to be smaller than $\approx 3\mu\text{m}\frac{r^3}{R^3}$, where $r$ is the typical dimension of the inner object, for the thermal motion to be confined near the center of the cavity. (This length scale is obtained by balancing $3 k_B T$ for room temperature with a rough estimate of the spring constant, $k\sim\frac{\hbar c}{R}\frac{r^3}{R^3}$.) To keep the two objects nearly concentric against the gravitational force, $R$ has to be smaller than $\approx1\mu\text{m}$ for the typical metal or liquid densities considered here.
A variety of applications may benefit from our analysis; in cancer therapy new treatments utilize nanocarriers that trap drug particles inside $R \approx .1\mu\text{m}$ polymer or lipid membrane shells~\cite{Peer07}, 
and molecular cages are proposed as containers for the storage of explosive chemicals \cite{Mal09}. Our results may guide the search for better materials and sizes of the enclosing cell.

Whether the center of the cavity is a point of stable or unstable equilibrium turns out to depend on whether the Casimir force is repulsive or attractive for two separated objects under those conditions. The direction of the torque, on the other hand, depends on the dielectric properties of the medium and the cavity walls in an unintuitive way, which cannot be predicted by the PSA or PFA, see \reffig{illus}. In particular, this behavior is not due to dispersion effects, which explain similar phenomena reported in Ref. \cite{Rodriguez08-2}. We calculate the torque on a small spheroid that is displaced from the center of a spherical cavity (\reffig{illus}~a)) or concentric with a slightly spheroidal cavity (\reffig{illus}~c)). In the former case, the orientation dependence manifests itself in $k$. In the latter case, $\mathcal{E}_0$ depends on the relative orientation of the spheroid inside the deformed cavity. (If the cavity is spherical, $\mathcal{E}_0$ is a finite constant that can be ignored.) By choosing appropriate materials, the inside object, e.g., a nanorod, can be made to align in different ways with the cavity shape, a situation which is reminiscent of a compass needle aligning with the magnetic field of the earth.

The starting point of the analysis is the scattering theory approach. The method is explained and derived in detail in Ref. \cite{Rahi09}, where a partial overview over its precursors, e.g. \cite{Emig07,Kenneth08,Maia_Neto08}, is provided. The Casimir energy
\begin{align}
\mathcal{E} &= \frac{\hbar c}{2 \pi } \int_0^{\infty} d\kappa \ln
\det (\mathcal{I} -  \mathcal{F}^{ii}_{O}  \mathcal{W}^{io}  \mathcal{F}^{ee}_{I}  \mathcal{V}^{io})
\label{eq:E}
\end{align}
is expressed in terms of the inner object's exterior $T$-matrix, $\mathcal{F}^{ee}_I$, and the outer object's interior $T$-matrix, $\mathcal{F}^{ii}_O$. The exterior $T$-matrix describes the scattering of regular wave functions to outgoing waves when the source lies at infinity. The interior $T$-matrix expresses the opposite, the amplitudes of the regular wave functions, which result from scattering outgoing waves from a source inside the object. The translation matrices $\mathcal{W}^{io}$ and $\mathcal{V}^{io}$ convert regular wave functions between the origins of the outer and the inner objects; they are related by complex transpose up to multiplication by $(-1)$ of some matrix elements.

With uniform, isotropic, and frequency-dependent permittivity $\ep_x(ic\kappa)$ and permeability $\mu_x(ic\kappa)$ functions ($x=I$: inner object; $x=O$: outer object; $x=M$: medium) the $T$-matrix of the sphere is diagonal. It is given by
\begin{align}
& \mathcal{F}^{ee}_{I,lmE,lmE}(ic\kappa)  = \mathcal{F}^{ee}_{I,lE}(\xi) = \nonumber \\
& -
\frac
{i_l(\xi) \partial_r (r i_l(z_I \xi))
- \tfrac{\epsilon_I}{\ep_M} i_l(z_I \xi) \partial_r (r i_l(\xi))}
{k_l(\xi) \partial_r (r i_l(z_I \xi))
- \tfrac{\epsilon_I}{\ep_M} i_l(z_I \xi) \partial_r (r k_l(\xi))}
\labeleqn{TEE}
\end{align}
for $E$ (electric) polarization and by the same expression with $\tfrac{\ep_I}{\ep_M}$ replaced by $\tfrac{\mu_I}{\mu_M}$ for $M$ (magnetic) polarization (not to be confused with subscript $M$ indicating the medium's response functions). In \refeqn{TEE} the frequency dependence of the response functions has been suppressed. The indices of refraction $n_x(ic\kappa)=\sqrt{\ep_x(ic\kappa)\mu_x(ic\kappa)}$ of the sphere and the medium appear in the ratio  $z_I(ic\kappa) = n_I(ic\kappa)/n_M(ic\kappa)$ and the argument $\xi=n_M(ic\kappa)\kappa r$. The first equality in \refeqn{TEE} defines an abbreviation for the $T$-matrix, in which the superfluous polarization and angular momentum $(l, m)$ indices are suppressed. The interior $T$-matrix of the spherical cavity is obtained from the exterior $T$-matrix of the sphere by inserting the outside object's radius and response function in place of those of the inside object and exchanging the modified spherical Bessel functions $i_l$ and $k_l$ everywhere.

However, the scattering approach is not limited to simple geometries. An array of techniques is available for calculating the scattering amplitudes of other shapes. We employ the perturbation approach to find the $T$-matrix of a deformed spherical cavity \cite{Yeh64,ErmaIII69} of radius $R+\delta(1-3/2\sin^2 \theta)$. The deformation, indicated in \reffig{illus} c), is chosen so that the volume is unchanged to first order in $\delta$. We find the $O(\delta)$ correction, $\mathcal{F}^{(1)}$, to the $T$-matrix in a perturbation series expansion, $\mathcal{F} = \mathcal{F}^{(0)}+\mathcal{F}^{(1)} + \cdots$, by matching the regular and outgoing fields according to the Maxwell boundary conditions along the deformed object's surface \cite{Millar73}. On the other hand, for a small object (compared to the wavelength of the radiation), we can approximate the $T$-matrix to lowest order in $\kappa$ using the static polarizability tensor, $\mathcal{F}^{ee}_{I,1mP,1m'P} = 2/3 (n_{M,0}\kappa)^3 \alpha_{mm'}^P + O(\kappa^5)$, where the subscript $0$ indicates the static $(ic\kappa=0)$ limit and $P$ is the polarization label. The $T$-matrix elements involving higher angular momenta $l>1$ are higher order in $\kappa$. For a small ellipsoid, in particular, the electric polarizability tensor $\boldsymbol{\alpha}^E$ is diagonal in a coordinate system aligned with the ellipsoid's body axes,
\begin{align}
\alpha^E_{ii} = \frac{V}{4\pi}\frac{\epsilon_{I,0}-\epsilon_{M,0}}{\epsilon_{M,0}+(\epsilon_{I,0}-\epsilon_{M,0})n_i},
\labeleqn{alphaE}
\end{align}
where $i\in \{x,y,z\}$ \cite{LandauLifshitz8}. The larger the semi-axis in direction $i$, the smaller the depolarization factor $n_i$ (not to be confused with the index of refraction). The entries of the magnetic polarizability tensor $\boldsymbol{\alpha}^M$ are obtained by exchanging $\mu_{x,0}$ for $\epsilon_{x,0}$ in \refeqn{alphaE}. In the small size limit the polarizability tensor of a perfect metal ellipsoid is obtained by taking both $\epsilon_{I,0}$ to infinity and $\mu_{I,0}$ to zero.


For simplicity we specialize to a spheroid, which has two equal semiaxes. We choose the semiaxes along $\mathbf{\hat{x}}$ and $\mathbf{\hat{y}}$ to be equal, therefore, $\alpha^P_{xx}=\alpha^P_{yy}=\alpha^P_{\perp\perp}$. The direction of displacement of the spheroid from the center of the cavity is the lab's $\mathbf{\hat{z}}$ axis. $\theta$ denotes the angle between the spheroid's and the lab's $\mathbf{\hat{z}}$ axes. For such a small spheroid inside a spherical cavity of radius $R$, the spring constant is obtained by expanding the log-determinant in \refeqn{E} to first order,
\begin{align}
k_{R\to\infty} & = \frac{\hbar c}{R^4 n_{M,0}}
\big[ \text{Tr } \boldsymbol{\alpha}^E f_1^E   \labeleqn{kasymp} \\ 
& +
\left(\alpha_{zz}^E-\alpha_{\perp\perp}^E\right) \tfrac{3\cos^2\theta-1}{2} f_2^E
+ E \to M \big], \nonumber
\end{align}
where the material dependent functions
\begin{align}
f_1^E & = \int_0^\infty \frac{\xi^5d\xi}{9\pi}\left[\mathcal{F}^{ii}_{O,1M}-2\mathcal{F}^{ii}_{O,1E}-\mathcal{F}^{ii}_{O,2E}\right] \, , \nonumber \\
f_2^E & = \int_0^\infty \frac{\xi^5d\xi}{9\pi}\left[\tfrac{4}{5}\mathcal{F}^{ii}_{O,1E}-\tfrac{1}{5}\mathcal{F}^{ii}_{O,2E}-\mathcal{F}^{ii}_{O,1M}\right]
\labeleqn{f}
\end{align}
express the rotation invariant and the orientation dependent parts of the energy, respectively. $f_1^M$ and $f_2^M$ are obtained from $f_1^E$ and $f_2^E$ by exchanging $E$ and $M$ everywhere.
This result is valid for asymptotically large $R$; it involves only the zero frequency ($ic\kappa=0$) response functions and the $l=1,2$ scattering amplitudes of the cavity walls in Eqs.~\refeq{kasymp} and~\refeq{f}.

\begin{figure}[ht]
\includegraphics[trim = 0mm 0mm 0mm 15mm,clip,width=8cm]{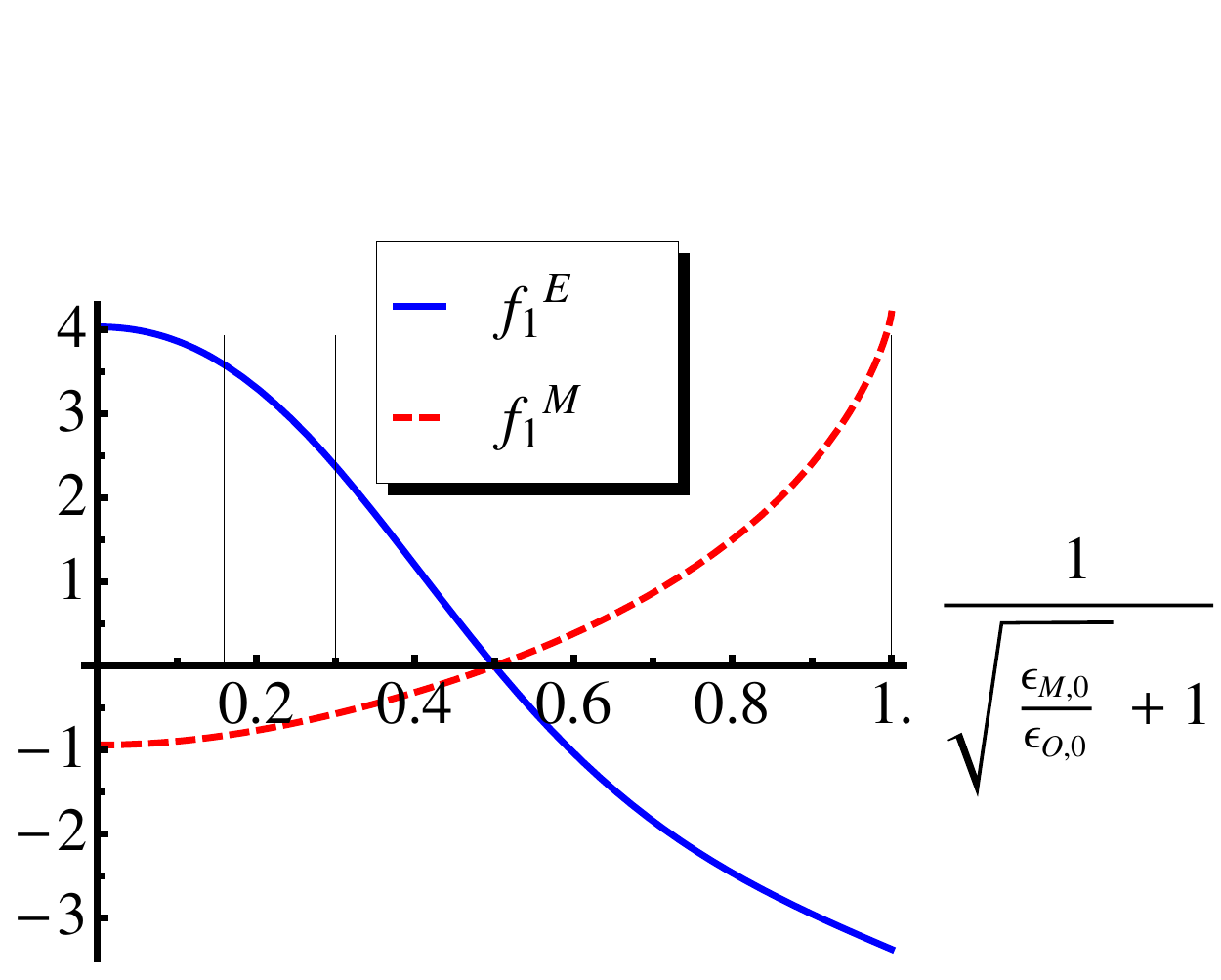}
\caption{(Color online) $f_1^E$ and $f_1^M$ describe the part of the spring constant $k_{R\to\infty}$, which is invariant under a rotation of the inside object. The vertical lines indicate the values pertaining to the configurations presented in Table~\ref{table1}, ethanol-vacuum (0.16), bromobenzene-vacuum (0.30), and gold cavity walls (1). In this plot, $\mu_{M,0}=\mu_{O,0}$.}
\label{fig:f1}
\end{figure}

The behavior of the functions $f_1^P$, depicted in \reffig{f1} for $\mu_{M,0}=\mu_{O,0}$, is as expected: $f_1^E$ is monotonic, positive when $\ep_{M,0}>\ep_{O,0}$, negative when $\ep_{M,0}<\ep_{O,0}$, and $f_1^M$ always has the opposite sign of $f_1^E$. When $f_1^E$ is positive, a small object with $\epsilon_{I,0}>\ep_{M,0}$ is levitated stably, when $f_1^E$ is negative, $\ep_{M,0}>\epsilon_{I,0}$ has to hold. Thus, stability occurs under the same conditions as repulsion for two objects outside of one another. The opposite sign of $f_1^M$ is expected from equivalent expressions for the two-infinite-slab geometry \cite{Kenneth02}. When the dielectric contrast between the medium and the outer sphere is taken to small or large limits, stability or instability is maximized.
To verify whether stability is observable for realistic materials and object sizes, we evaluate \refeqn{E} numerically for a sphere of radius $r$ inside a spherical cavity of radius $R$ filled with various liquids.~%
\footnote{For the permittivity function of gold we use $\epsilon(ic\kappa) = 1+\frac{\omega_p^2}{c\kappa(c\kappa+\gamma)}$, where $\omega_p=1.14 \cdot 10^{16} \text{Hz}$ and $\gamma=9.27 \cdot 10^{13} \text{Hz}$ \cite{Pirozhenko06} as used in Ref. \cite{Munday09}. For the other materials we use an oscillator model, $\ep(i c \kappa) = 1 + \sum_{n=1}^N \frac{C_n}{1+\left(c \kappa/\omega_n\right)^2}$, with $[\omega_n]_{n=1,2}=[6.6,114]\cdot10^{14}\text{Hz}$ and $[C_n]_{n=1,2}=[23.84,0.852]$ for ethanol \cite{Milling96} as used in Ref. \cite{Rodriguez08-2}, with $[\omega_n]_{n=1,2}=[5.47,128.6]\cdot10^{14}\text{Hz}$ and $[C_n]_{n=1,2}=[2.967,1.335]$ for bromobenzene \cite{Milling96} as used in Ref. \cite{Munday09}, and with $[\omega_n]_{n=1-4}=[0.867,1.508,2.026,203.4]\cdot10^{14}\text{Hz}$ and $[C_n]_{n=1-4}=[0.829,0.095,0.798,1.098]$ for silica ($\text{SiO}_2$) \cite{Bergstroem97}. A discussion of permittivity models is given, for example, in Ref. \cite{Parsegian05}.}
The coefficients $k$, $k_4$, and $k_6$ in the series expansion in \refeqn{Eexp} are listed in Table \ref{table1}. For comparison, the asymptotic result $k_{R\to\infty}$ is also included. If both inside and outside object are spherically symmetric, the series in \refeqn{Eexp} does not contain terms $\sim\tfrac{a^n}{R^n}$ with $n$ odd.

\begin{table}[h]
\caption{$k$, $k_{R\to\infty}$, $k_4$, and $k_6$ are listed for various combinations of materials for the case depicted in \reffig{illus} b). The dimensionless numbers in the table have to be multiplied by $\tfrac{\hbar c}{R}$. $R$ is given in microns $[\mu\text{m}]$. $k_{R\to\infty}$ depends on $R$ only through the ratios $\tfrac{\hbar c}{R}$ and $\tfrac{r}{R}$, so its numerical prefactor is the same for all $R$. The highest cutoff used was $l_\text{max}=30$.}
\label{table1}
\begin{tabular}{| l | l | l | l | l | l | l |}
\hline
Inside-Medium-Outside &
\multicolumn{1}{|c|}{$R$} &
\multicolumn{1}{|c|}{$r/R$} &
\multicolumn{1}{|c|}{$k$} &
\multicolumn{1}{|c|}{$k_{R\to\infty}$} &
\multicolumn{1}{|c|}{$k_4$} &
\multicolumn{1}{|c|}{$k_6$}
\\
\hline
Gold-Bromobenzene- & 0.1 & 1/4 & 4.0e-2 & 5.4e-2 & 1.2   & 8.2e1 \\
Vacuum             &     & 3/4 & 2.2e1  & 1.4    & 4.2e3 & 1.8e6 \\
                   & 1.0 & 1/4 & 6.9e-2 & 5.4e-2 & 2.6   & 2.0e2 \\
                   &     & 3/4 & 7.0e1  & 1.4    & 1.8e4 & 1.0e7 \\
\hline
Gold-Ethanol-      & 0.1 & 1/4 & 5.0e-2 & 3.7e-2 & 1.6   & 1.0e2 \\
Vacuum             &     & 3/4 & 2.7e1  & 9.9e-1 & 5.2e3 & 2.2e6 \\
                   & 1.0 & 1/4 & 4.5e-2 & 3.7e-2 & 1.7   & 1.4e2 \\
                   &     & 3/4 & 6.0e1  & 9.9e-1 & 1.8e4 & 1.2e7 \\
\hline
Silica-Bromobenzene- & 0.1 & 1/4 & 6.1e-3 & 7.1e-3  & 1.9e-1 & 1.3e1 \\
Gold                 &     & 3/4 & 4.2    & 1.9e-1  & 8.6e2  & 4.0e5 \\
                     & 1.0 & 1/4 & 1.8e-2 & 7.1e-3  & 7.3e-1 & 6.0e1 \\
                     &     & 3/4 & 2.3e1  & 1.9e-1  & 5.7e3  & 3.1e6 \\
\hline
Silica-Ethanol-    & 0.1 & 1/4 & 1.5e-2 & 1.2e-2 & 4.6e-1 & 3.1e1 \\
Gold               &     & 3/4 & 1.0e1  & 3.3e-1 & 1.9e3  & 8.4e5 \\
                   & 1.0 & 1/4 & 1.9e-2 & 1.2e-2 & 8.2e-1 & 7.2e1 \\
                   &     & 3/4 & 4.1e1  & 3.3e-1 & 1.2e4  & 7.6e6 \\
\hline
\end{tabular}
\end{table}

The three materials were chosen so that the sequence of permittivities $\ep_I$, $\ep_M$, $\ep_O$ either increases or decreases for the imaginary frequencies that dominate in \refeqn{E}. Contrary to the prediction of the PSA and PFA the force is not symmetric with respect to exchange of the inner and outer permittivities. In the same medium, a high dielectric sphere is held more stably in the center of a cavity with low dielectric walls than a low dielectric sphere inside a cavity with high dielectric walls.

The asymptotic result $k_{R\to\infty}$ yields a good approximation of $k$ for $\tfrac{r}{R}=\tfrac{1}{4}$. But from \refeqn{kasymp} one would expect $k_{R\to\infty}$ to grow linearly with the volume of the inner sphere, since polarizability is proportional to the volume, see \refeqn{alphaE}. In fact, $k$ for $\tfrac{r}{R}=\tfrac{3}{4}$ is about $1000$ times larger than for $\tfrac{r}{R}=\tfrac{1}{4}$, instead of just $27$ times. This means that for a gold sphere in a liquid drop with $\tfrac{r}{R}=\tfrac{3}{4}$ and $R=1\mu\text{m}$ at room temperature, indeed, the Casimir spring holds the particle near the center effectively, $\sqrt{\langle a^2 \rangle}/R < 0.1$. 

\begin{figure}[ht]
\includegraphics[trim = 0mm 0mm 0mm 30mm,clip,width=8cm]{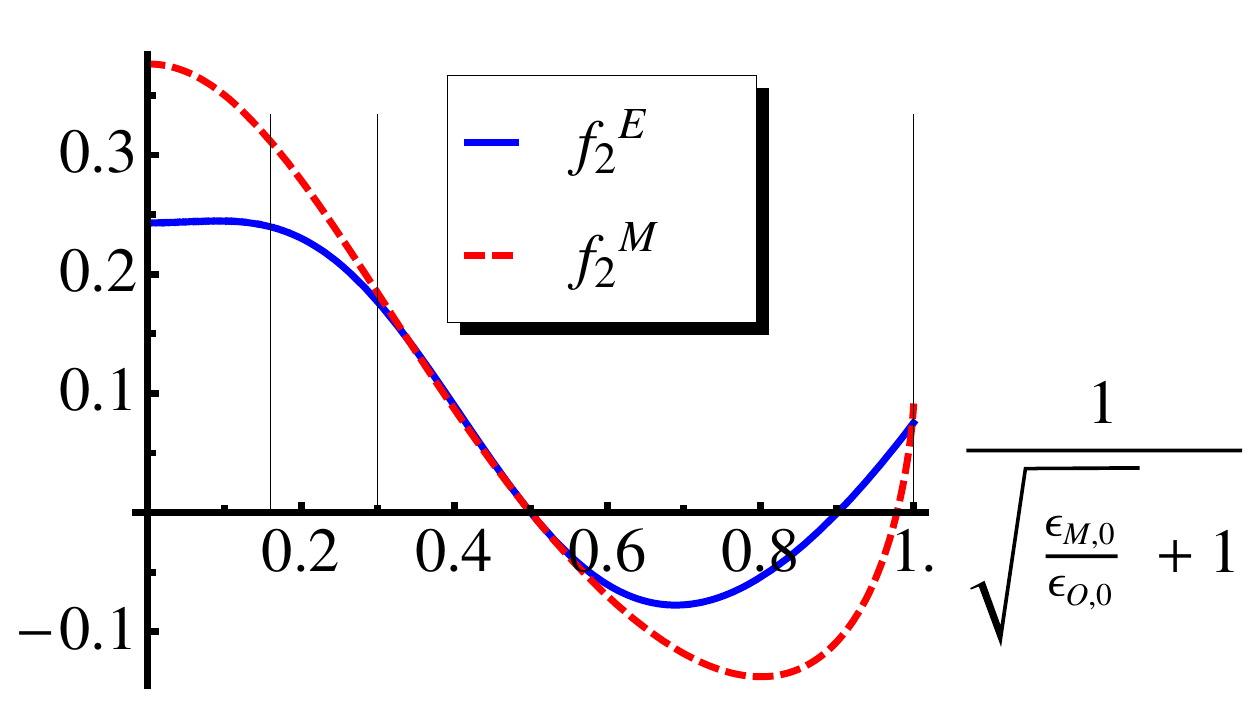}
\caption{(Color online) $f_2^E$ and $f_2^M$ describe the part of the spring constant $k_{R\to\infty}$, which changes with the orientation of the inside spheroid. In this plot, $\mu_{M,0}=\mu_{O,0}$.}
\label{fig:f2}
\end{figure}

Compared to the stability conditions studied thus far, the orientation dependence of the energy is more varied. $f_2^E$ and $f_2^M$, plotted in \reffig{f2}, have the same sign for most ratios of medium to outside permittivities, unlike $f_1^E$ and $f_1^M$, which always have opposite signs. In these ranges of values, the contributions to the torque from electric and magnetic polarizability are opposite for a small perfect metal spheroid, for which $\epsilon_{I,0}>\epsilon_{M,0}$ and $\mu_{I,0}< \mu_{M,0}$. Unlike $f^E_1$ and $f^M_1$, also, $f^E_2$ and $f_2^M$ change sign again at $\frac{\epsilon_{O,0}}{\epsilon_{M,0}}\approx 80$ and at $\frac{\epsilon_{O,0}}{\epsilon_{M,0}}\approx 2000$, respectively. So, while the direction of the total force and the stability of the equilibrium can be predicted based on the relative magnitudes of the permittivities, the torque cannot be. The second sign change of $f_2^E$ and $f_2^M$ does not agree with the PSA or PFA. Furthermore, it raises the question whether calculations of the Casimir torque for infinite conductivity metals are `universal' in the sense that they produce the correct qualitative results for real materials.

\begin{figure}[ht]
\includegraphics[trim = 0mm 0mm 0mm 25mm,clip,width=8cm]{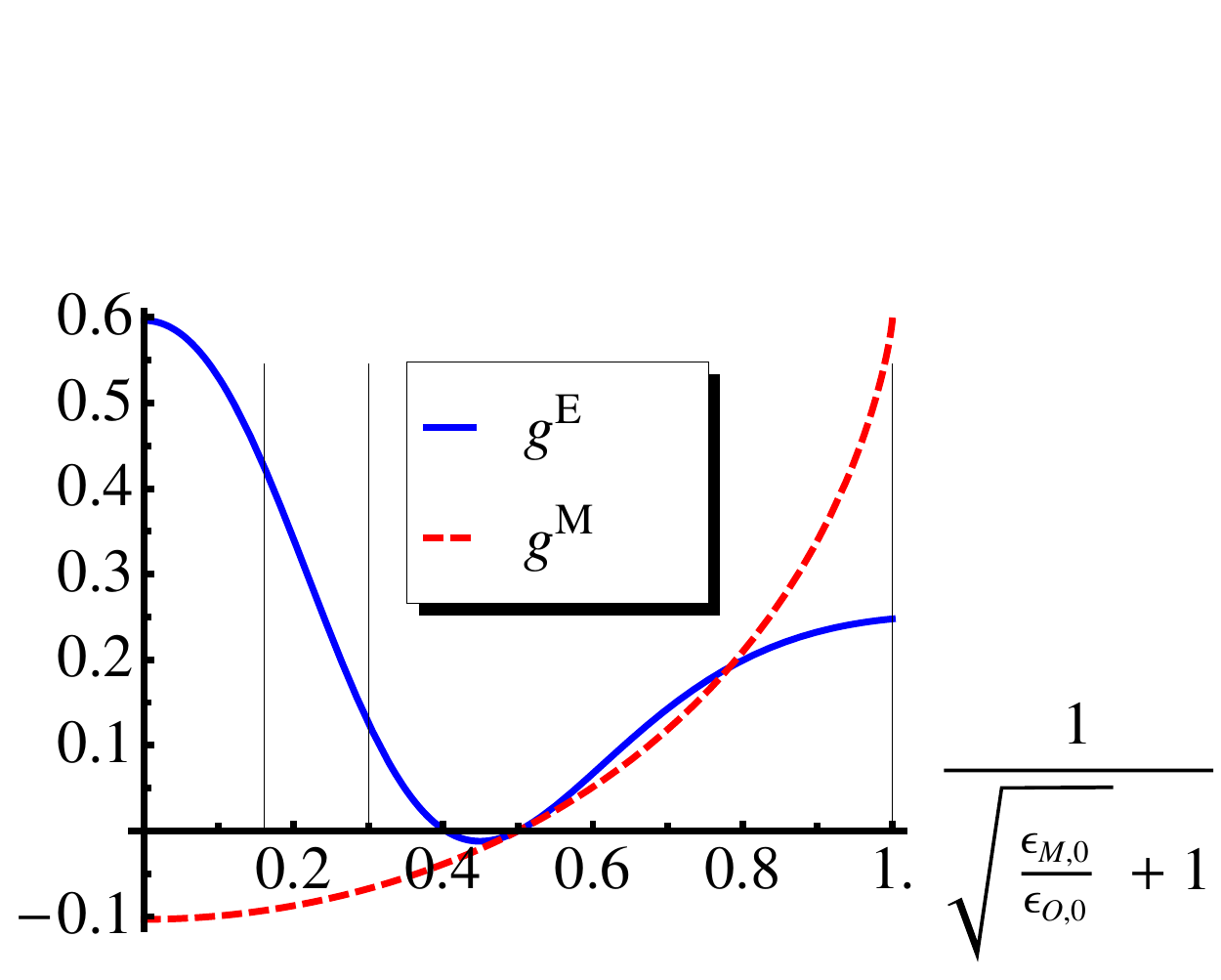}
\caption{(Color online) $g^E$ and $g^M$ describe the dependence of the energy $\mathcal{E}_0$ on the relative orientation of the inside spheroid and the deformed cavity walls.}
\label{fig:g}
\end{figure}

The orientation dependent part of the energy for the configurations discussed thus far vanishes, of course, when the small spheroid is in the center of the spherical cavity. If the cavity is slightly deformed, however, the energy, $\mathcal{E}_0$, depends on the relative orientation of the spheroid and the cavity. We deform the spherical cavity as described earlier and obtain to first order in $\delta/R$,
\begin{align}
\mathcal{E}_0 = \frac{\hbar c \, \cos^2 \theta}{R^4 n_{M,0}} \frac{\delta}{R} 
\left[ \left(\alpha_{zz}^E-\alpha_{\perp\perp}^E\right) g^E + E\to M \right],
\end{align}
where the orientation-independent part of the energy has been dropped. For $\mu_{O,0}=\mu_{M,0}$, $g^E$ and $g^M$ are given by
\begin{align}
g^E & =\int_0^\infty \frac{\xi^4d\xi}{10\pi}
\tfrac{i_1 k_0 + i_0 k_1}{\left(\tilde{k}_0 i_1 + \tilde{k}_1 i_0 z_O + i_1 \tilde{k}_1/(z_O\xi)(1-z_O^2)\right)^2}
\nonumber \\
& \times
(z_O^2-1)
\left(4 \tilde{k}^2_1/\xi^2-(\tilde{k}_0+\tilde{k}_1/(z_O\xi))^2\right), \nonumber \\
g^M & =\int_0^\infty \frac{\xi^4d\xi}{10\pi}
\tfrac{i_1 k_0 + i_0 k_1}{\left(\tilde{k}_1 i_0 + \tilde{k}_0 i_1 z_O\right)^2}
(z_O^2-1)
\tilde{k}^2_1,
\end{align}
where the arguments of the modified spherical Bessel functions and of $z_O(0)$ are suppressed. $i_l$ and $k_l$ are functions of $\xi$, and $\tilde{k}_l$ stands for $k_l(z_O(0) \xi)$, where $z_O(ic\kappa)=n_O(ic\kappa)/n_M(ic\kappa)$ is the ratio of the permittivities of the cavity walls and the medium. The material dependent functions $g^E$ and $g^M$ are plotted in \reffig{g}.

Again, $g^E$ has the same sign for $\tfrac{\epsilon_{O,0}}{\epsilon_{M,0}}\to 0$ (left in \reffig{g}) and $\tfrac{\epsilon_{O,0}}{\epsilon_{M,0}}\to \infty$ (right). In addition to the root at $\tfrac{\epsilon_{O,0}}{\epsilon_{M,0}}=1$, $g^E$ also vanishes at $\tfrac{\epsilon_{O,0}}{\epsilon_{M,0}}\approx 0.46$. 

The rich orientation dependence of the energy is expected to collapse as the size of the inside object grows to fill the cavity and the PFA becomes applicable. Based on the stability analysis for finite inside spheres, though, we expect the asymptotic results, $f_2^P$ and $g^P$, to predict the orientation dependence for reasonably small inside spheroids.

In reality, various corrections to the idealized shapes have to be taken into account. A drop of liquid, which is placed on a surface, is influenced by gravity and interactions with the substrate. An asymmetric inner object, or one that is displaced from the center, deforms the shape of the droplet additionally by causing uneven Casimir stresses.

This work was supported by the NSF through grant DMR-08-03315 (SJR). We thank T. Emig, N. Graham, R. Jaffe, and M. Kardar for fruitful discussions.

\end{document}